\documentclass[aps,%
reprint,
prl,
groupedaddress]{revtex4-2}
\usepackage{xcolor}
\usepackage{float}
\usepackage{amsmath,amsfonts,amssymb,epsfig,graphicx}
\usepackage{slashed} 
\usepackage{hyperref}
\usepackage[normalem]{ulem}

\newcommand{\PKp}{\ensuremath{\mathrm{K^+}}}
\newcommand{\PKm}{\ensuremath{\mathrm{K^-}}}
\newcommand{\PKl}{\ensuremath{\mathrm{K^0_L}}}
\newcommand{\PKs}{\ensuremath{\mathrm{K^0_S}}}

\newcommand{\MeV}{\,\text{MeV}}

\newcommand{\fbinv}{\mbox{\ensuremath{~\mathrm{fb^{-1}}}}}

\graphicspath{{./pic/}}
\begin{document}

\title{New Methods to Achieve Meson, Muon and Gamma Light Sources Through Asymmetric Electron Positron Collisions}

\author{Dawei \surname{Fu}}
\email[]{fudw@pku.edu.cn}
\affiliation{State Key Laboratory of Nuclear Physics and Technology, School of Physics, Peking University, Beijing, 100871, China}

\author{Alim \surname{Ruzi}}
\email[]{alim.ruzi@pku.edu.cn}
\affiliation{State Key Laboratory of Nuclear Physics and Technology, School of Physics, Peking University, Beijing, 100871, China}

\author{Meng \surname{Lu}}
\email[]{meng.lu@cern.ch}
\affiliation{School of Physics, Sun Yat-Sen University, Guangzhou 510275, China}

\author{Qiang \surname{Li}}
\email[]{qliphy0@pku.edu.cn}
\affiliation{State Key Laboratory of Nuclear Physics and Technology, School of Physics, Peking University, Beijing, 100871, China}

\begin{abstract}
We propose methods to produce gamma light source, muon, and energetic meson beams such as charged and neutral Kaons, which are boosted to be collimated and with relatively long life time. The first type of methods is based on asymmetric electron positron collisions with a center of mass energy of, e.g., 1020 MeV, and Kaons can be produced at a rate of $10^{4-5}/s$. The electron and positron beams are either asymmetric in energy, e.g., 10 GeV electron beam with 26 MeV positron beam, or asymmetric in space, e.g., 10 GeV electron and positron beams collisions with a angle around 0.05 radius. Such proposals should be able to be achieved with a reasonable budget. The other type of method is relying on TeV positron on target experiment, where Kaon beams can be achieved at around $10^{7}$ per bunch crossing. Such Kaon beams are clean with small contamination, and can have great physics potential on, e.g., hyperon searches through Kaon nuclei collision, Kaon rare decay measurement, and Kaon proton or Kaon lepton collisions. The same technique can also be extended to other final states such as pions and tau leptons. 
\end{abstract}

\maketitle

Novel collision methods and rich phenomena are crucial to keeping high-energy collision physics more robust and attractive~\cite{Lu:2022ibc}. Recent years have witnessed vast development towards next generation high energy colliders, including various proposals on Higgs factory~\cite{EuropeanStrategy}, revived interest in Muon collider~\cite{Aime:2022flm}, etc. In this letter, we are interested in possible source and usage of energetic collimated meson beams. We start from the case of Kaons, which contain rich physics as explained below.

Strange Hadron Spectroscopy is an interesting topic and relevant proposal has been made e.g. with the KLong Facility at Jefferson Lab~\cite{KLF:2020gai}, where KL beam can provide unique data for spectroscopy. The idea is to create a secondary beam of neutral kaon with  a flux on the order of $1\times 10^4$ \PKl\ per second.

Rare Kaon decays experiments, such as NA62~\cite{NA62:2021zjw}, provide sensitive probes of the short distance physics.  Among the most promising modes are the $\PKp\rightarrow\pi^+\nu\bar\nu $ and $\PKl\rightarrow\pi^0\nu\bar\nu$ modes due to their theoretical cleanness  and the dedicated experiments.  $\PKl\rightarrow\pi^0\ell^+\ell^-$ and $\PKl\rightarrow\mu^+\mu^-$ decays might also be interesting probes~\cite{Stamou:2011aj}. One of the most significant importance of the Kaon decay is that it might shed light on the search of the new physics, by observing the CP-violated decays $\PKl\rightarrow\pi^0\nu\bar\nu$ because of the purely imaginary hadronic matrix element~\cite{NA62KLEVER:2022nea}.

Rare Kaon decays also offer a new window on another important area of physics by producing neutrino and antineutrino pairs with well known flux. These neutrinos can serve as the source for Neutrino Factory~\cite{IDS-NF:2011swj}, which can be a complement to Muon-Based Neutrino Study~\cite{Huber:2014nga}. Kaon decay into same sign dimuon can also serve as probes of the effective Majorana mass of $M_{\mu\mu}$. 

Kaon-related experiments are typically based on proton target~\cite{NA62:2021zjw}, or tertiary beam from electron beam converted to photons and then to Kaons~\cite{KLF:2020gai}. 
On the other hand, Kaons can be copiously produced at electron-positron collision process $e^+e^-\rightarrow \phi(1020)\rightarrow \PKp\PKm,\PKl\PKs$, especially at the resonant energy 1020 \MeV. 

\begin{figure}
    \centering
    \includegraphics[width=.9\columnwidth]{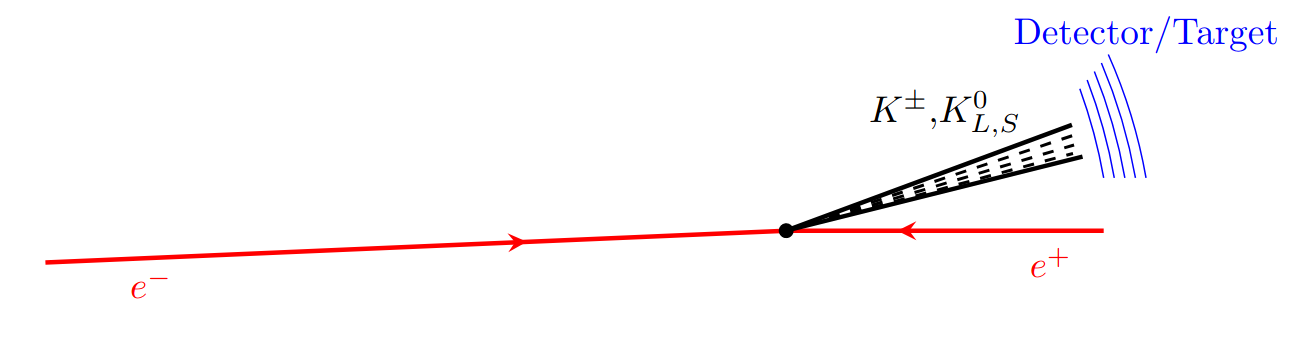}
    \caption{Kaons produced at a rate of $10^{4-5}/s$ from asymmetric electron-positron collisions at e.g. 1020 MeV.}
    \label{fig:eeK}
\end{figure}

\begin{figure}[H]
    \centering
    \includegraphics[width=7cm, height=5.3cm]{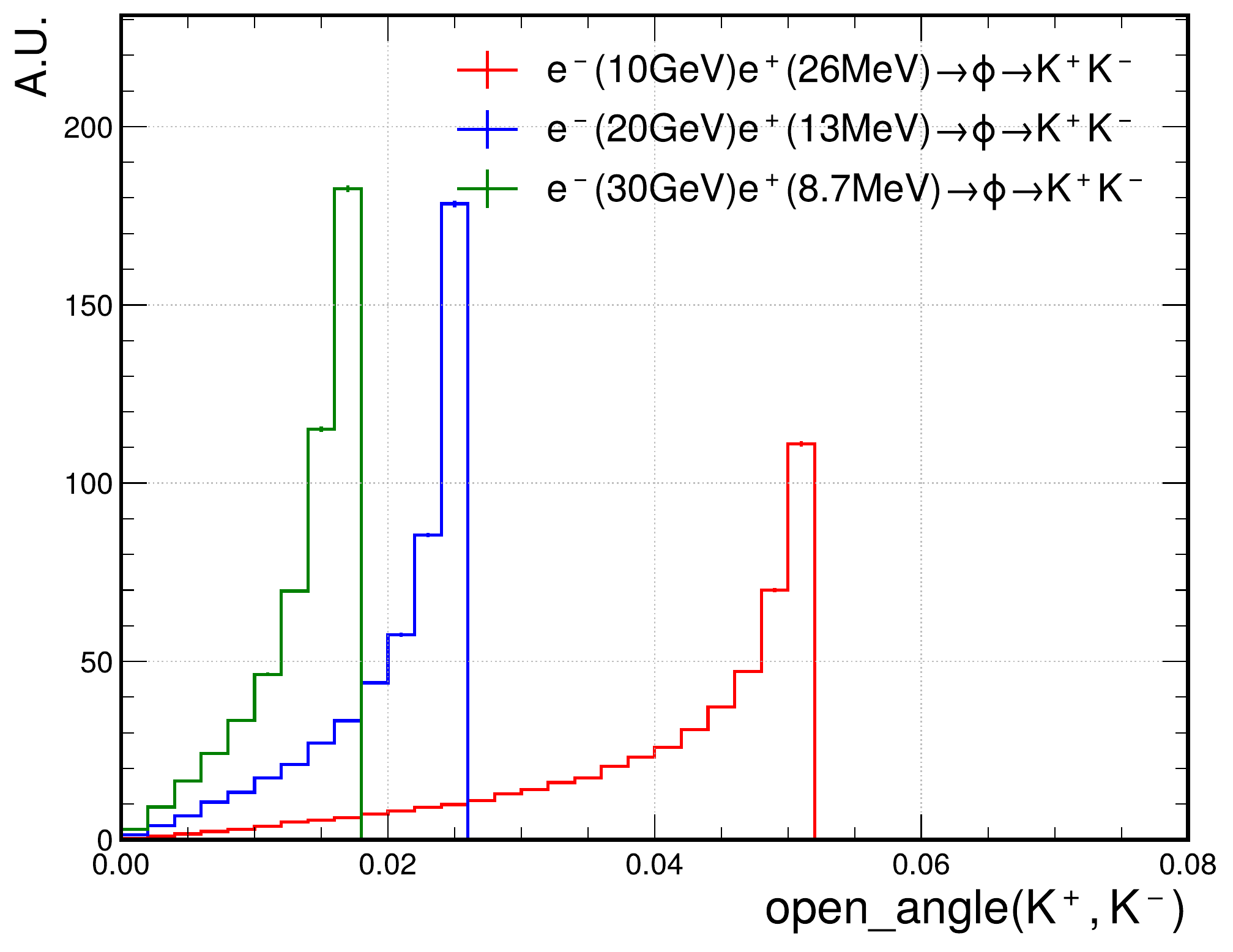}
    \caption{Distributions of the opening angle between the final state Kaon pairs from $e^+e^-\rightarrow \phi(1020)\rightarrow \PKp\PKm,\PKl\PKs$, for an asymmetric collision profile with higher energy electron and low energy positron. The center of mass energy is set to be 1020 \MeV. }
    \label{fig:angle}
\end{figure}

Accordingly, we propose here two novel methods to produce energetic meson beams such as charged and neutral Kaons, which are boosted to be collimated with relatively long life time. The first method, as shown in Fig.~\ref{fig:eeK}, is based on highly asymmetric electron-positron collisions  with a center-of-mass energy of,  e.g., 1020 MeV, and Kaons can be produced at a rate of $10^{4-5}/s$. The other method is relying on TeV-scale positron on fixed-target experiment, where Kaon beam can achieve around $10^{7}$ per bunch crossing. Such Kaon beams can have great physics potential, such as hyperon searches via Kaon-proton collision,  Kaon rare decay measurement, and Kaon-Kaon or Kaon-lepton collisions.

1) Electron beam with medium energy of 10 GeV or above and positron beam with low energy around 30 MeV can annihilate into Kaon pairs via $\phi(1020)$ resonances. Due to large Lorentz boost, ${\rm K}^{\pm}$ and ${\rm K}^0_{L}$ can fly over a long distance of $\sim10^2$ meters. The cross section is around 4 micro-barns, while the instantaneous luminosity for electron-positron collider can typically reach $ {\cal L}=10^{34}$ cm$^{-2}$s$^{-1}$ or above. Thus the Kaon production rate can be estimated to be $10^{4-5}$ per second, which can already be comparable with or even exceeds the KLong Facility~\cite{KLF:2020gai}.

Fig.~\ref{fig:angle} shows the distributions of the opening angle between the final state Kaon pairs from $e^+e^-\rightarrow \phi(1020)\rightarrow \PKp\PKm,\PKl\PKs$, for an asymmetric collision profile with higher energy electron and low energy positron. The center-of-mass energy is set to be 1020 \MeV. Kaons are observed to be relatively collimated, with a typical opening angle below 0.05 radius. For the case of 20 GeV electron colliding with 13 MeV positron beams, the angle can be mostly smaller than around 0.025 radius. The beam spread size counts around 0.25 meters for a distance of 10 meters from the interaction point, thus an economical detector should work here with a reasonable material budget. Furthermore, differently charged Kaons can be separated from neutral Kaons with a sweep magnet~\cite{KLF:2020gai}, and both can be exploited for rich physics studies.

\begin{figure}
    \centering
    \includegraphics[width=.9\columnwidth]{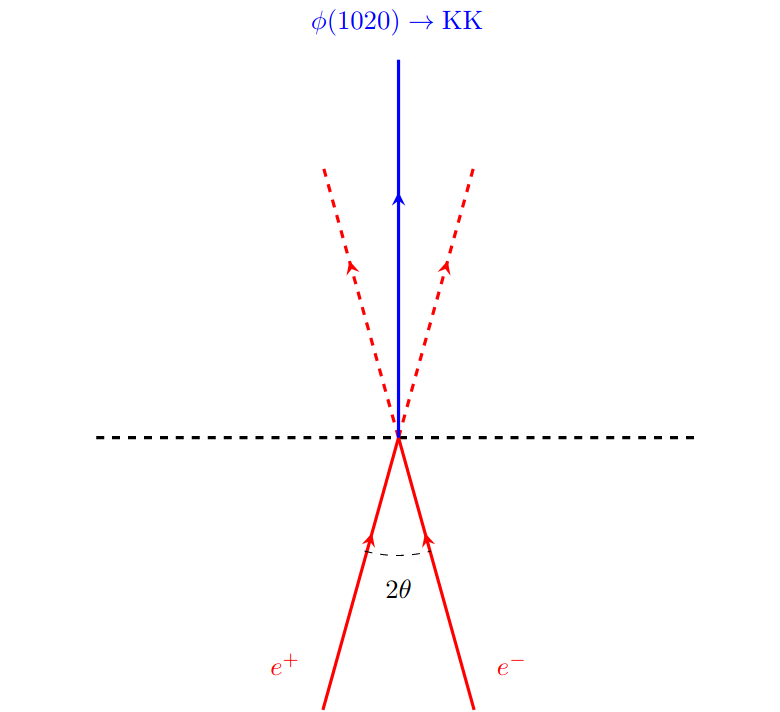}
    \caption{Collisions between electron and positron beams separated by an angle $2\theta$. The two beams are symmetric in energy, while asymmetric in space. For electron and positron beams with energy of 5 (or 10) GeV, to achieve $\phi(1020)$ resonant productions, the angle is $\theta\sim 0.102\,(\text{or }0.051)$ radius.}
    \label{fig:assym}
\end{figure}

While preparing for this draft, we found in the 1990s similar proposals were proposed on an asymmetric phi factory~\cite{Eberhard:1992kr,Cline:1994ir}, which includes details about beam design and luminosity estimation. Either linac or ring deign for electron or positron beams can be exploited. One of the key challenges towards very asymmetric collisions is the beam beam tune shift which may degrade machine luminosity significantly~\cite{Eberhard:1992kr,Cline:1994ir}. The degradation may be mitigated in the linac-linac collision mode~\cite{Eberhard:1992kr,Cline:1994ir}, especially in referring to vast development and progresses in the designs of the ILC and CLIC. Moreover, we propose a new collision pattern as shown in Fig.~\ref{fig:assym}, where the two beams are symmetric in energy, while asymmetric in space, i.e. the two beams are separated by an angle of $2\theta$. For electron and positron beams with energy of 5 GeV (or 10 GeV), to achieve $\phi(1020)$ resonant productions, the angle is $\theta\sim 0.102\,(\text{or } 0.051)$ radius. Notice that KEKB has a similar size of crossing angle around  22 mrad at the interaction point, and achieved a luminosity of $2\times 10^{34}$ cm$^{-2}$s$^{-1}$ with the crab Cavities~\cite{Funakoshi:2014yfa}. An estimation has also been provided in ref.~\cite{Batygin2002} on luminosity for two comoving beams intersecting each other at a small angle, which finds that it is possible to reach similar luminosity as for symmetric nominal collisions although it is necessary to provide long interaction region with short bunches.

As for Kaon's rare decay measurement, one of the principal disadvantages of the NA62-like experiment lies in the difficulty of separating the kaons from other charged hadrons in the beam. This can be easily overcome in our case, as the Kaon beams here are from lepton collisions with much cleaner environment. With $10^7$ seconds of operations, i.e. around one year, $10^{11-12}$ Kaons can be produced and decay in the forward detector symbolized as in Fig.~\ref{fig:eeK} (the detector design can refer to NA62's). The collected data lead to competitive sensitivity on $\PKp\rightarrow\pi^+\nu\bar\nu $.  Crucially, unlike NA62, our proposal can provide simultaneously similar sensitivities on properties measurement for both $\PKm$ and $\PKp$, and neutral and charged Kaons.  Moreover, as our Kaon beam energy is typically at the 10 GeV level, the decay length will be shorter than NA62 which operates with a Kaon beam energy of around 75 GeV. 

As for Strange Hadron Spectroscopy, the Kaon beams can be shed on targets, interfacing with a GlueX-like detector, symbolized as in Fig.~\ref{fig:eeK}. Charged Kaons may be swept away with specific magnet~\cite{KLF:2020gai}. Thanks to the higher energy and clean Kaon beams proposed here, one can probe Hyperons or other strange hadrons at higher mass region compared to the KLong Facility~\cite{KLF:2020gai,Dobbs:2022agy}, through processes such as ${\rm K}_{\rm L} {\rm N} \to (\Lambda^*,\Sigma^*) \to {\rm K}^+ \Xi^**$.

2) On the other hand, positron on target has been exploited for high-quality muon beam production~\cite{Alesini:2019tlf}. To produce $\phi(1020)$ and Kaons similarly, the positron energy should be around 1.02 TeV, which could be possible from high energy muon decays or from ILC/CLIC beam techniques. The number of Kaon pairs produced per positron bunch on target $n(\mu^+ \mu^-)_{max}\approx n^+\times 10^{-4\,{\rm to}\,-5}$.

Similar studies can surely be done similarly as discussed above for Kaon spectrum and rare decay measurements or searches. However, now with more energetic and collimated Kaon beams, we are more interested in head-on Kaon collisions with a lepton or proton beam at the TeV energy scale. Although the Kaon beam density is lower than the lepton or proton beam by 3 or 4 orders of magnitudes, an integrated luminosity of 1-10 \fbinv should still be achievable for such a novel meson collider in 10 years' time scale.

\begin{figure}
    \centering
    \includegraphics[width=.9\columnwidth]{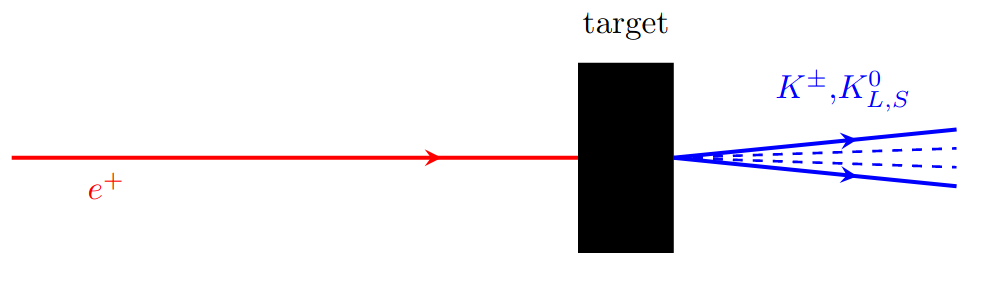}
    \caption{Collimated Kaons produced from TeV scale positron dumped at target.}
    \label{fig:eon}
\end{figure}

The Kaon beam is more continuously distributed from the parental electron-positron collisions, it would be possible to squeeze kaons within a ring and produce bunches for further collision usage. As Kaon contains a large portion of strange quark at high momentum fraction, collisions of Kaon with electron or muon can directly produce leptoquarks copiously, which couples to a lepton and a quark at the same time and are one of the most popular candidates to explain the Lepton Flavour Universality anomalies~\cite{LHCb:2021trn}. Collisions between Kaon and Proton beams may also shed light on the strange quark distribution functions or spin information in hadrons. Finally, it may be possible to probe Higgs couplings with strange quarks through ${\rm s\bar{s}}\rightarrow {\rm H}$ at appropriate collision energy.

In Summary, We propose here two types of novel methods to produce energetic meson beams such as charged and neutral Kaons, which are boosted to be collimated with a relatively long lifetime. 
The first type of method is based on asymmetric electron-positron collisions with a center-of-mass energy of, e.g., 1020 MeV, and Kaons can be produced at a rate of $10^{4-5}/s$. The electron and positron beams are either asymmetric in energy, e.g., 10 GeV electron beam with 26 MeV positron beam,  or asymmetric in space, e.g., 10 GeV electron and positron beams collisions with an angle around 0.05 radius. Such proposals should be able to be achieved with a reasonable budget.
The other type of method is relying on TeV-scale positron on the fixed-target experiment, where Kaon beams can achieve around $10^{7}$ per bunch crossing. Such Kaon beams are clean with small contamination and have great physics potential on, e.g., hyperon searches through Kaon-nuclei collision, Kaon rare decay measurement, and Kaon-proton or Kaon-lepton collisions. 

\begin{figure}
    \centering
    \includegraphics[width=.9\columnwidth]{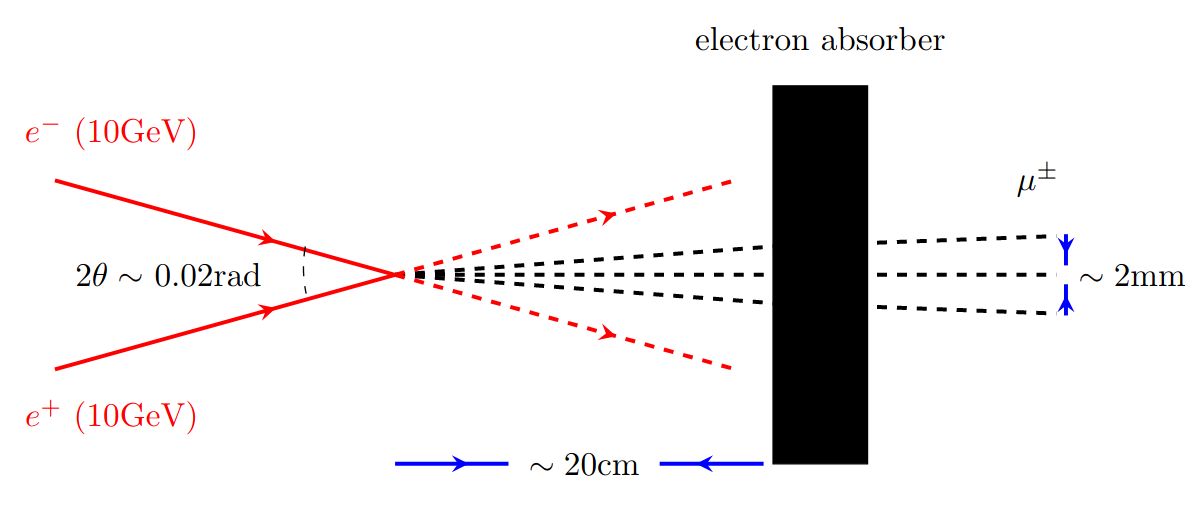}
    \caption{Collimated muon beams produced from collisions between electron and positron beams both with the energy of 10 GeV and separated by an angle $2\theta\sim 0.02$ rad. Downstream there inserts an electron absorber, and the output muon beam size is around 2mm level.}
    \label{fig:eemm}
\end{figure}

The above-mentioned methods are based on asymmetric electron-positron collisions and can be extended further to generate collimated beams for other final state particles similarly, for example, pion beams can be produced via $e^+e^-\rightarrow \omega(782)$, and collimated muon or tau leptons can be produced through $e^+e^-\rightarrow $ $\mu^+\mu^-$, $\tau^+\tau^-$ (see Fig.~\ref{fig:eemm}). For the case of $\omega(782)$ decaying into photons, it is possible to achieve 10 GeV level gamma light source with the rate of $10^{4-5}$/s, which is comparable or surpass the power of current GeV gamma light factories~\cite{Muramatsu:2021bpl}. These beams will surely benefit particle physics and material science.


\appendix
\begin{acknowledgments}
This work is supported in part by the National Natural Science Foundation of China under Grants No. 12150005, No. 12075004, and No. 12061141002, by MOST under grant No. 2018YFA0403900.
\end{acknowledgments}

\bibliographystyle{ieeetr}
\bibliography{h}
\end{document}